\documentclass[final,5p,times,twocolumn]{article}

\usepackage[utf8]{inputenc}
\usepackage{natbib}
\usepackage{graphicx}
\usepackage[margin=0.5in]{geometry}

\usepackage{lineno,hyperref}
\modulolinenumbers[5]

\usepackage{epstopdf}
\usepackage{verbatim}
\usepackage{mathrsfs}
\usepackage{array}

\newcolumntype{H}{>{\setbox0=\hbox\bgroup}c<{\egroup}@{}}

\begin{document}

\title{A simple direct empirical observation of systematic bias of the redshift as a distance indicator}

\date{}

\author{Lior Shamir\footnote{lshamir@mtu.edu}  \\ Kansas State University \\ 1701 Platt St \\ Manhattan, KS 66506, USA}

\maketitle

\abstract{
Recent puzzling observations such as the $H_0$ tension, large-scale anisotropies, and massive disk galaxies at high redshifts have been challenging the standard cosmological model. While one possible explanation is that the standard model is incomplete, other theories are based on the contention that the redshift model as a distance indicator might be biased. These theories can explain the recent observations, but they are challenged by the absence of a direct empirical reproducible observation that the redshift model can indeed be inconsistent. Here I describe a simple experiment that shows that the spectra of galaxies depend on their rotational velocity relative to the rotational velocity of the Milky Way. Moreover, it shows that the redshift of galaxies that rotate in the opposite direction relative to the Milky Way is significantly smaller compared to the redshift of galaxies that rotate in the same direction relative to the Milky Way ($p<0.006$). Three different datasets are used independently, each one was prepared in a different manner, all of them show similar redshift bias. A fourth dataset of galaxies from the Southern Galactic pole was also analyzed, and shows similar results. All four datasets are publicly available. While a maximum average z difference of $\sim$0.012 observed with galaxies of relatively low redshift (z$<$0.25) is not extreme, the bias is consistent, and can potentially lead to explanations to puzzling observations such as the $H_0$ tension.
}



\section{Introduction}
\label{introduction}

Recent observations have shown unexplained tensions and anomalies at cosmological scales. For instance, the $H_0$ determined by the Cosmic Microwave Background (CMB) radiation is different from the $H_0$ determined by using Ia supernovae and the redshift of their host galaxies  \citep{wu2017sample,mortsell2018does,bolejko2018emerging,davis2019can,pandey2020model,camarena2020local,di2021realm,riess2022comprehensive}. The relatively new JWST provides unprecedented imaging power, showing massive disk galaxies with mature stellar populations at unexpectedly high redshifts that can be as high as 15 \citep{whitler2023ages}. In fact, the presence of large disk galaxies at unexpectedly high redshifts was reported also before JWST saw first light \citep{neeleman2020cold}. The existence of these galaxies is unexpected, as pre-JWST analyses predicted that such galaxies do not exist according to $\Lambda$CDM cosmology \citep{cowley2018predictions}, and therefore were not expected to be observed.

These unexpected observations challenge our understanding of the Universe. If the common distance indicators are complete and fully accurate, the standard cosmological theories are incomplete. Similarly, if the standard cosmological model is complete, it is not possible that the distance indicators as currently used are fully accurate. Therefore, assuming that all distance indicators are reliable and consistent, explaining these observations might reinforce certain modifications of some of the foundations of cosmology. But in addition to theories that shift from the standard cosmological model, other theories are based on the contention that the redshift as a distance indicator at cosmological scales might be biased \citep{crawford1999curvature,pletchermature,gupta2023jwst,lee2023cosmological,seshavatharam2023rotating,seshavatharam2023understanding,lovyagin2022cosmological}. 
While the assumption that the redshift might be biased or inconsistent can explain these observations without modifying the standard model, there is currently no clear reproducible empirical evidence that the redshift might indeed be biased.

The redshift of a luminous moving object is determined by the linear component of the Doppler shift effect. But because galaxies have rotational velocity in addition to their linear velocity, their redshift can also be affected by the rotational velocity, as the rotational velocity of a luminous object also leads to a Doppler shift effect \citep{marrucci2013spinning,lavery2014observation,liu2019experimental}.  

Since the rotational velocity of a galaxy is far smaller than its linear velocity relative to Earth, the rotational velocity component of the Doppler shift is often ignored  when determining the distance of a galaxy. But while the Doppler shift effect driven by the rotational velocity of the galaxy is expected to be subtle, that has not yet been tested. It should also be mentioned that the physics of galaxy rotation is one of the most mysterious observations, and its nature cannot be explained unless making assumptions such as dark matter \citep{zwicky1937masses,oort1940some,rubin1983rotation,el2020aedge}, modified Newtonian dynamics (MOND) \citep{milgrom1983modification,milgrom2007mond,de1998testing,sanders1998virial,sanders2002modified,swaters2010testing,sanders2012ngc,iocco2015testing,diaz2018emergent,falcon2021large}, or other theories \citep{sanders1990mass,capozziello2012dark,chadwick2013gravitational,farnes2018unifying,rivera2020alternative,nagao2020galactic,blake2021relativistic,gomel2021effects,skordis2021new,larin2022towards}.  But despite over a century of research, there is still no single clear proven explanation to the physics of galaxy rotation  \citep{sanders1990mass,mannheim2006alternatives,kroupa2012dark,kroupa2012failures,kroupa2015galaxies,arun2017dark,akerib2017,bertone2018new,aprile2018,skordis2019gravitational,sivaram2020mond,hofmeister2020debate,byrd2021spiral,haslbauer2022high,haslbauer2022has}, and that phenomenon is still not fully understood. 

The purpose of the simple experiment described in this paper is to test the impact of the rotational velocity component of galaxies on the Doppler shift effect, and consequently on the redshift as a distance indicator. Section~\ref{data} describes the data used in the experiment, Section~\ref{results} provides the results of the analysis, Section~\ref{other_datasets} compares the results shown in Section~\ref{results} to several other datasets collected by different methods and different telescope systems, and Section~\ref{explanation} discusses possible explanations in the light of recently observed anisotropies and current cosmological theories.

\section{Data}
\label{data}

The experiment is based on one primary dataset, and two additional independent datasets of galaxies from the Northern Galactic pole to which the results are compared. Each of these datasets was prepared in a different manner. A fourth dataset of galaxies from the Southern Galactic pole is also used for comparison, as will be discussed in Section~\ref{southern_pole}. 

The primary dataset contains  SDSS DR8 galaxies with spectra sorted by their direction of rotation, as explained and used in \citep{shamir2020patterns}. Instead of using galaxies in the entire SDSS footprint, this experiment is focused on galaxies that rotate in the same direction relative to the Milky Way, and galaxies that rotate in the opposite direction relative to the Milky Way. Therefore, only galaxies that are close to the Galactic pole are used, and the field is limited to the $20\times20$ degrees centered at the Northern Galactic pole. {Obviously, because the Milky Way rotates around its own pole, galaxies that are close to the galactic pole rotate either in the same direction relative to the galactic pole, or in the opposite direction relative to the galactic pole.} The analysis included objects with spectra in SDSS DR8 that have r magnitude of less than 19 and Petrosian radius of at least 5.5''. The redshift of the galaxies in that initial set was limited to z$<$0.3, and the redshift error smaller than $10^{-4}$. That selection eliminated the possible effect of bad redshift values, that in some cases can be very high and skew the dataset. The initial set of galaxies that meet these criteria in that field contained 52,328.


The process by which the galaxies were sorted by their direction of rotation is explained in detail in \citep{shamir2020patterns}, and is similar to the process of annotating galaxies imaged by other telescopes \citep{shamir2016asymmetry,shamir2020asymmetry,shamir2022analysis,shamir2022using,shamir2022asymmetry,mcadam2023reanalysis,shamir2022possible2}. In summary, the annotation is done by using the Ganalyzer algorithm \citep{shamir2011ganalyzer}, where each galaxy image is transformed into its radial intensity plot such that the value of the pixel at Cartesian coordinates $(\theta,r)$ in the radial intensity plot is the median value of the 5$\times$5 pixels at coordinates $(O_x+\sin(\theta) \cdot r,O_y-\cos(\theta)\cdot r)$ in the original galaxy image, where {\it r} is the radial distance measured in percentage of the galaxy radius, $\theta$ is the polar angle in degrees relative to the galaxy center, and $(O_x,O_y)$ is the coordinates of the galaxy center. A peak detection algorithm is then applied to the rows in the radial intensity plot, and the direction of the peaks determines the direction of the curves of the galaxy arms. 

\begin{figure*}
\centering
\includegraphics[scale=0.75]{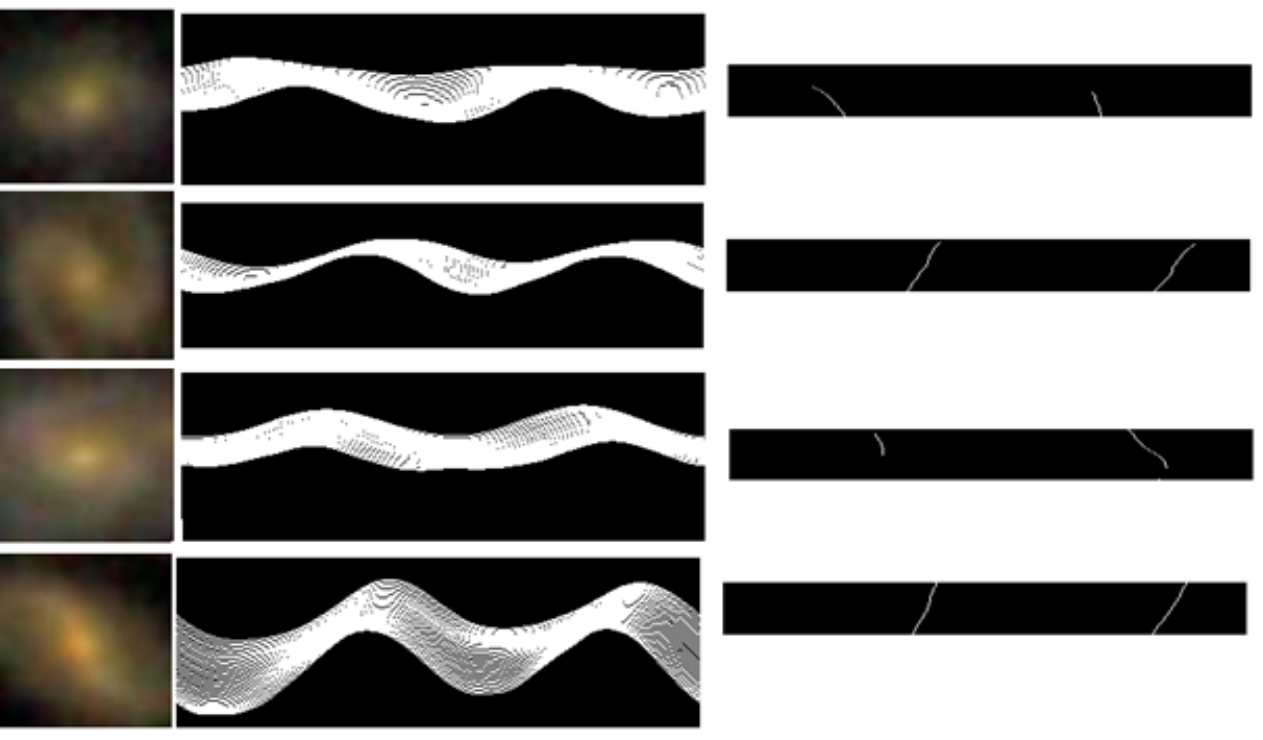}
\caption{Examples of original galaxy images (left), the radial intensity plot transformations (center), and the peaks detected in the the radial intensity plot lines (right).}
\label{radial_intensity_plot}
\end{figure*}

Figure~\ref{radial_intensity_plot} displays examples of the original galaxy images, their radial intensity plots, and the detected peaks. The direction of the curves of the arms is determined by the sign of the slope. Based on previous experiments \citep{shamir2020patterns,shamir2020asymmetry,shamir2022analysis,shamir2022using,shamir2022asymmetry,mcadam2023reanalysis,shamir2022possible2}, to avoid incorrect annotations, a direction was determined only for galaxies that had at least 30 peaks are identified in the radial intensity plot. If less than 30 peaks are identified the galaxy is not used, as its direction of rotation cannot be identified. The algorithm is described with experimental results in \citep{shamir2011ganalyzer}, as well as \citep{shamir2020asymmetry,shamir2022analysis,shamir2022using,shamir2022asymmetry,mcadam2023reanalysis}.

The primary advantage of the algorithm is that its simple ``mechanical'' nature makes it fully symmetric. Experiments of mirroring the galaxy images lead to identical inverse results compared to when using the original images \citep{shamir2016asymmetry,shamir2020asymmetry,shamir2022analysis,shamir2022using,shamir2022asymmetry,mcadam2023reanalysis}. 

\begin{figure}
\centering
\includegraphics[scale=0.7]{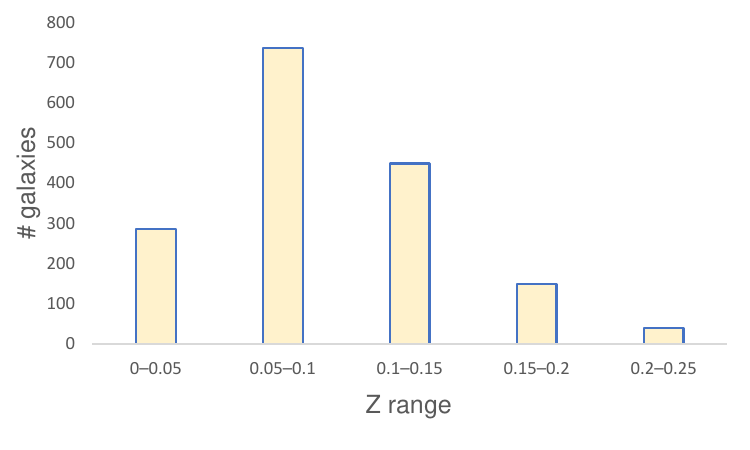}
\caption{The redshift distribution of the galaxies in the dataset.}
\label{redshift_range}
\end{figure}

After applying the algorithm to the galaxy images, the final dataset included 1,642 galaxies with identifiable directions of rotation, such that 817 galaxies rotate clockwise, and 825 galaxies rotate counterclockwise. Applying the algorithm to the mirrored images led to an identical inverse dataset. Testing a random subset of 200 galaxies showed that all galaxies were annotated correctly. Figure~\ref{redshift_range} shows the redshift distribution of the galaxies. The dataset is available at \url{https://people.cs.ksu.edu/~lshamir/data/zdifference/}. In addition to this dataset, two other previous public datasets were used, as will be described in Section~\ref{other_datasets}.

\section{Results}
\label{results}

Table~\ref{main_dataset} shows the redshift differences in the 20$\times$20 degree field centered at the Northern Galactic pole, as well as the smaller 10$\times$10 degree field. The errors are the standard error of the mean, computed by $\frac{\sigma_z}{\sqrt{N}}$, where $N$ is the number of galaxies. The mean redshift of the galaxies in the dataset described in Section~\ref{data} that rotate in the same direction relative to the Milky Way (MW) is 0.09545$\pm$0.0017, while the mean redshift of the galaxies in the same field that rotate in the opposite direction relative to the Milky Way (OMW) is 0.08895$\pm$0.0016. That shows a $\Delta z$ of $\sim$0.0065$\pm$0.0023 between galaxies that rotate in the same direction relative to the Milky Way and galaxies that rotate in the opposite direction relative to the Milky Way. 

By applying a simple Student t-test, the two-tailed probability that the two means are different by mere chance is ($p\simeq$0.0058). 
To verify the statistical significance, a simulation test was also performed. The simulation randomly selected a set of 817 galaxies, and another set of 825 galaxies regardless of their spin direction. That was repeated 100,000 times, and in each run the mean redshift of the galaxies in the first set were compared to the mean redshift of the galaxies in the second set. In 614 of the runs the difference between the mean redshifts of the two sets was greater than 0.0016. That provide a probability of 0.0061, which is similar to the two-tailed t-test probability. The similarity between the probability of the t-test and the probability of the simulation is not surprising, but provides consistency and ensures that the t-test P-values are not driven by a certain unusual distribution of the redshift values.

If the observed difference in redshift is driven by the rotational velocity of the observed galaxy relative to the rotational velocity of the Milky Way, the difference should increase when the observed galaxies are closer to the Galactic pole. As Table~\ref{main_dataset} shows, $\Delta z$ indeed increases in the 10$\times$10 field. Despite the lower number of galaxies the difference is still statistically significant.

\begin{table*}
\caption{The mean redshift difference of galaxies in 20$\times$20 degree field centered at the Galactic pole and the 10$\times$10 degree field centered at the Galactic pole. The galaxies are separated into galaxies that rotate in the same direction relative to the Milky Way (clockwise to an Earth-base observer) and galaxies that rotate in the opposite direction relative to the Milky Way (OMW). The P values are the two-tailed P values determined by the standard Student t-test. The errors are the standard error of the mean.}   
\label{main_dataset}
\centering
\begin{tabular}{lcccccc}
\hline
Field  ($^o$)      & \# MW  & \# OMW  & $Z_{mw}$  & $Z_{omw}$  & $\Delta$z   & t-test P \\
\hline
10$\times$10     &   204   &  202  & 0.0996$\pm$0.0036   &  0.08774$\pm$0.0036  & 0.01185$\pm$0.005 & 0.02 \\  
20$\times$20     &   817   &  825  & 0.09545$\pm$0.0017  &  0.08895$\pm$0.0016 &  0.0065$\pm$0.0023 & 0.0058 \\  
\hline
\end{tabular}
\end{table*}

\begin{figure}
\centering
\includegraphics[scale=0.7]{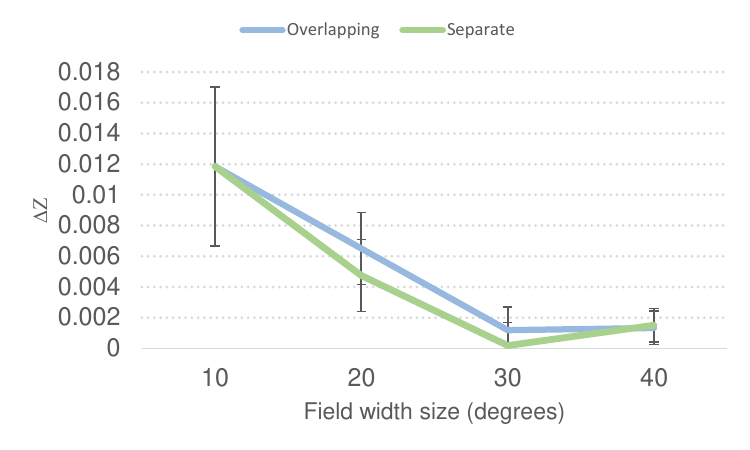}
\caption{The $\Delta z$ when the size of the field changes. The analysis was done such that the larger field contains also the galaxies of the smaller field inside it (blue), and also when the galaxies in the smaller field are excluded so that the two fields are orthogonal, and do not have overlapping galaxies (green).}
\label{distances} 
\end{figure}

If the redshift difference peaks at the Northern Galactic pole, it is expected that galaxies that are on or close to the galactic pole would show higher redshift difference, while when using galaxies that are more distant from the Galactic pole the redshift difference $\Delta z$ would decrease. Figure~\ref{distances} shows the change in $\Delta z$ when the size of the field centered at the Galactic pole changes.

As the figure shows, the $\Delta z$ decreases as the field gets larger. That can be explained by the fact that when the field gets larger, it includes more galaxies that are more distant from the Galactic pole. While it does not fully prove a link to the rotational velocity of these galaxies relative to the Milky Way, that observation is in agreement with the contention that the redshift difference is linked to the rotational velocity of the galaxies relative to the rotational velocity of the Milky Way. The figure includes two graphs. The first shows all galaxies inside the field. For instance, when the field size is 20$\times20$ degrees it also includes the galaxies inside the 10$\times$10 degree field centered at the Galactic pole. The other analysis excludes overlapping galaxies, so that a galaxy can only be used in one field. That is, the galaxies in the 20$\times$20 degree field centered at the Galactic pole excludes the galaxies in the 10$\times$10 degree field. That provides analysis with independent sets of galaxies that do not overlap. The analysis shows a small $\Delta z$ of $\sim$0.002 also when the dataset includes galaxies that are more distant from the Galactic pole, but it is smaller compared to the $\Delta z$ observed when the set of galaxies is limited to galaxies closer to the Galactic pole.

To test a possible change in different redshift ranges, the galaxies in the 20$\times$20 field centered at the Galactic pole were separated into galaxies with redshift lower than 0.1 and redshift greater than 0.1. Using 0.1 for separating the dataset into lower redshift and higher redshift galaxies provided two sets, one with 1,012 galaxies with z$<$0.1, and another dataset of 630 galaxies with z$>$0.1. Table~\ref{z_ranges} shows the differences in redshift when the galaxies are separated to the two different redshift ranges. The $\Delta z$ of the galaxies where $z<0.1$ is 0.0028$\pm$0.0013, while when $z>0.1$ the $\Delta z$ is 0.0053$\pm$0.0028. The t-test probability that the difference occurs by chance is P$\simeq$0.51, and therefore not statistically significant. While not statistically significant, the dataset used here might not be sufficiently large to provide statistically significant conclusions regarding the $\Delta z$ in the different redshift ranges, if such indeed exist.  

\begin{table*}
\caption{The mean redshift difference of galaxies with z$<$0.1 and z$>$0.1 that rotate in the same direction as the Milky Way or in the opposite direction relative to the Milky Way.}   
\label{z_ranges}
\centering
\begin{tabular}{lccccc}
\hline
z range            & \# MW  & \# OMW  & $Z_{mw}$  & $Z_{omw}$  & $\Delta$z   \\
\hline
z$<$0.1     &   491   &  521  & 0.0629$\pm$0.001   &  0.0601$\pm$0.001  & 0.0028$\pm$0.0013  \\  
z$>$0.1     &   326   &  304  & 0.1441$\pm$0.002  &  0.1388$\pm$0.002 &  0.0053$\pm$0.0028  \\  
\hline
\end{tabular}
\end{table*}



  

Table~\ref{spec_dif} shows the differences between the flux on the different filters, taken from the specObjAll table in SDSS DR8. The spectrum flux difference shows a consistent difference of $\sim$10\% across the different filters. Unlike the redshift, the differences in the flux of the specific filters are not statistically significant, and therefore a definite conclusion about the flux differences cannot be made.

\begin{table*}
\caption{Flux in different filter galaxies that rotate in the same direction relative to the Milky Way and galaxies that rotate in the opposite direction relative to the Milky Way. The t-test P values are the two-tailed P value.}   
\label{spec_dif}
\centering
\scriptsize
\begin{tabular}{lcccccc}
\hline
Band            & MW           & OMW  & $\Delta$   & t-test P \\
\hline
spectroFlux\_g  &   25.969$\pm$0.8669        &  28.554$\pm$1.0918     & -2.585  &  0.063 \\  
spectroFlux\_r   &   53.2433$\pm$1.765        & 58.6214$\pm$2.3422    &  -5.378  &  0.066 \\  
spectroFlux\_i   &   77.4189$\pm$2.513        &  85.0868$\pm$3.407     &  -7.667  &  0.067 \\  

\hline
\end{tabular}
\end{table*}

\section{Comparison to other datasets}
\label{other_datasets}

The annotation algorithm used to sort the galaxies by their direction of rotation as discussed in Section~\ref{data} is simple and symmetric, and there is no known bias that can prefer the redshift of a certain set of galaxies as annotated by the algorithm. Also, experimenting with the same images when the images were mirrored leads to inverse results, as also shown in detail in \citep{shamir2016asymmetry,shamir2020asymmetry,shamir2021large,shamir2022analysis,shamir2022using,shamir2022asymmetry,mcadam2023reanalysis}. To further test for a possible impact of unknown or unexpected biases in the annotation process, two additional annotation methods were used to test whether these algorithms provide different results.

\subsection{Comparison to annotations by {\it Galaxy Zoo}}
\label{galaxy_zoo}

The first annotation method that was used is the crowdsourcing-based {\it Galaxy Zoo 1} \citep{lintott2008galaxy}. In {\it Galaxy Zoo}, anonymous volunteers used a web-based interface to sort galaxy images by their direction of rotation. After several years of work by over 100,000 volunteers, a relatively large set of over $8\cdot10^5$ galaxies were annotated. One of the downsides of {\it Galaxy Zoo} was that in the vast majority of the cases the volunteers who annotated the galaxies made conflicting annotations, and the disagreement between the annotators makes it difficult to use the majority of the galaxies. Another substantial downside is that the annotations were subjected to the bias of the human perception, which is very difficult to model and fully understand, challenging the reliability of the annotations as a tool for primary science. Despite these known weaknesses, there is no known human perceptual bias that would associate galaxies with lower redshift to a certain direction of rotation. Therefore, although {\it Galaxy Zoo} might not necessarily be considered a complete tool when used as the sole dataset, comparing to {\it Galaxy Zoo} can provide an indication of whether a different annotation method leads to different results shown in Section~\ref{results}.

Because the annotations of the volunteers often disagree with each other, {\it Galaxy Zoo} defined the ``superclean'' criterion as galaxies that 95\% of the human annotators agree on the annotation. That is, if 95\% of the annotations or more are for a galaxy that rotates clockwise, the annotation is considered ``superclean''. While these annotations are normally correct, only 1.7\% of the galaxies annotated by {\it Galaxy Zoo 1} meet that criterion. Out of the 667,944 galaxies in the specZoo table in SDSS DR8, just 324 galaxies meet that criterion and are also inside the 20$\times$20 degree field centered at the Northern Galactic pole.

The mean z of the {\it Galaxy Zoo 1} galaxies that rotate in the same direction relative to the Milky Way in that field is 0.073834$\pm$0.0041. The mean z of the galaxies that rotate in the opposite direction relative to the Milky Way is 0.068292$\pm$0.00348. That shows a $\Delta z$ of 0.00554, which is similar in both direction and magnitude to the $\Delta z$ of 0.0065 observe with the dataset described in Section~\ref{data}. The one-tailed P value of that difference to occur by mere chance is 0.15. That is not statistically significant, and that can be attributed to the small size of the dataset, but the similar $\Delta z$ in both direction and magnitude shows consistency between the annotation methods. From the 324 galaxies annotated by Galaxy Zoo, 263 were also included in the dataset described in Section~\ref{data}. The value of the comparison is therefore not by analyzing a new dataset, but by using a different annotation method that is independent of the method used in Section~\ref{data}, and therefore not subjected to the same possible unknown or unexpected biases in that method if such exist. An analysis of a completely different dataset acquired by a different telescope system and does not have any overlap with the SDSS dataset will be described in Section~\ref{southern_pole}.

\subsection{Comparison to annotations by {\it SpArcFiRe}}
\label{sparcfire}

Another dataset that is used is the dataset of SDSS galaxies annotated by the {\it SpArcFiRe} (Scalable Automated Detection of Spiral Galaxy Arm) algorithm \citep{Davis_2014,hayes2017nature}. {\it SpArcFiRe} is implemented by an open source software \footnote{https://github.com/waynebhayes/SpArcFiRe}, and the method is described in detail in \citep{Davis_2014}. In summary, the algorithm first identifies arm segments in the galaxy image, and then fits these segments to logarithmic spiral arcs to determine the direction of rotation based on the curves of the arms. One of the advantages of {\it SpArcFiRe} is that it is not based on data-driven machine learning or deep learning approaches that are difficult to analyze, and is therefore not subjected to the complex biases that are often very difficult to notice \citep{dhar2022systematic}. The downside of {\it SpArcFiRe} is that it has an annotation error of about 15\% \citep{mcadam2023reanalysis}. More importantly, since {\it SpArcFiRe} is a relatively sophisticated algorithm, it is more difficult to ensure that it is completely symmetric, and in some seldom cases a mirrored galaxy image is not annotated as rotating to the opposite direction compared to the original image. That characteristic of the algorithm is discussed in the appendix of \citep{hayes2017nature}. That weakness of the algorithm can be addressed by repeating the analysis twice, such that in the first experiment the original images are used, and in the second experiment the mirrored images are used. Then, the results of the two experiments can be compared. While that practice might not be ideal, it can be used to compare the results to the results shown in Section~\ref{results}.

The dataset used here is the dataset of spiral galaxies annotated {\it SpArcFiRe} used in \citep{mcadam2023reanalysis}, which is a reproduction of the experiment described in \citep{hayes2017nature}. As explained in \citep{hayes2017nature}, the set of galaxies is the same set of galaxies selected for annotation in \citep{lintott2008galaxy}, although the manual annotations are not used in any form in the analysis. The dataset is available at \url{https://people.cs.ksu.edu/~lshamir/data/sparcfire}. More details about the dataset are available in \citep{mcadam2023reanalysis}. 

The dataset was prepared with the original images, and then again with the mirrored galaxy images. The dataset prepared with the original images contains 138,940 galaxies, and the dataset prepared with the mirrored images contains 139,852 galaxies. That difference is expected due to the fact that the {\it SpArcFiRe} method is not fully symmetric, as explained in the appendix of \citep{hayes2017nature}. All of the galaxies used in the experiment have spectra, and therefore can be used to compare the reshift. As before, galaxies with redshift greater than 0.3 or redshift error greater than $10^{-4}$ were ignored. Table~\ref{table_sparcfire} shows the mean redshift in the 10$\times$10 field centered at the Northern Galactic pole and in the 20$\times$20 field, for both the original images and the mirrored images.

\begin{table*}
\caption{The mean redshift of galaxies annotated by the {\it SpArcFiRe} algorithm. The table shows results when using the original images, as well as the results when the algorithm is applied to the mirrored images, leading to inverse $\Delta z$. The t-test P values are the one-tailed P value.}   
\label{table_sparcfire}
\centering
\begin{tabular}{lcccccc}
\hline
Field    ($^o$)   & \# MW  & \# OMW  & $Z_{mw}$  & $Z_{omw}$  & $\Delta$z   & t-test P \\
\hline
Original  10$\times$10     &   710       &  732  &  0.07197$\pm$0.0015  &  0.06234$\pm$0.0014 & 0.00963$\pm$0.002 & $<$0.0001 \\  
Mirrored  10$\times$10   &   728       &   709  &  0.06375$\pm$0.0014  & 0.07191$\pm$0.0014  & -0.00816$\pm$0.002 & $<$0.0001 \\  
Original  20$\times$20     &   2903   &  2976  & 0.07285$\pm$0.0007  &  0.071164$\pm$0.0007 &  0.001686$\pm$0.0009 & 0.04 \\  
Mirrored  20$\times$20   &   3003   &   2914  & 0.07113$\pm$0.0007  &  0.07271$\pm$0.0007  &  -0.00158$\pm$0.0009 & 0.05 \\  
\hline
\end{tabular}
\end{table*}

As the table shows, both the original images and the mirrored images show consistent results. These results are also consistent with the results shown in Section~\ref{results}. The $\Delta z$ is lower than the $\Delta z$ observed with the dataset used in Section~\ref{results}, and that could be due to the certain error rate of the annotations made by the {\it SpArcFiRe} algorithm, which is expected to weaken the signal as also shown quantitatively in Section 7.1 in \citep{mcadam2023asymmetry}.  To test for the effect of the annotation error, the dataset described in Section~\ref{data} was used such that 15\% of the galaxies were assigned intentionally with the wrong spin direction. That reduced the $\Delta z$ from 0.0065 as shown in Section~\ref{results} to 0.0032$\pm$0.0023, which is closer to the $\Delta z$ value shown in Table~\ref{table_sparcfire}, and within 1$\sigma$ statistical fluctuation from it. That also shows that the magnitude of the $\Delta z$ depends on the accuracy of the annotation, which is another indication to the link between the spin direction of the galaxies and the $\Delta z$.

\subsection{Comparison to galaxies from the Southern Galactic pole}
\label{southern_pole}

The data used in the experiments described above was all taken from the Northern hemisphere, and the galaxies it contains are around the Northern Galactic pole. To verify the observed redshift difference, it is also required to test if it exists in the Southern Galactic pole as well. If the observed difference in redshift is also observed in the Southern Galactic pole, it can provide an indication that it indeed could be related to the Galactic pole. Since the three experiments above all used data collected by SDSS, using a different telescope can show that the difference is not driven by some unknown or unexpected anomaly in a specific telescope system. Also, a dataset of galaxies from the Southern Galactic pole will have no overlap with the three datasets of SDSS galaxies used above.

The set of galaxies used for the analysis are galaxies imaged by DECam used in \citep{shamir2021large} that had spectroscopic redshift through the Set of Identifications Measurements and Bibliography for Astronomical Data (SIMBAD) database \citep{wenger2000simbad}. As explained in \citep{shamir2021large}, DECam galaxy images were acquired through the API of the DESI Legacy Survey server. The initial set of galaxies contained all objects in the South bricks of DESI Legacy Survey Data Release 8 that had g magnitude of less than 19.5 and identified as de Vaucouleurs $r^{1/4}$ profiles (“DEV”), exponential disks (“EXP”), or round exponential galaxies (‘’REX”).

The galaxy images were then annotation by the Ganalyzer algorithm as described in Section~\ref{data}, and also in \citep{shamir2021large}. Unlike the other methods used in Sections~\ref{galaxy_zoo} and~\ref{sparcfire}, Ganalyzer provides a high level of accuracy in the annotations. Also, DESI Legacy Survey does not have datasets of galaxies annotated through as used in Section~\ref{galaxy_zoo}. 

\begin{figure}
\centering
\includegraphics[scale=0.6]{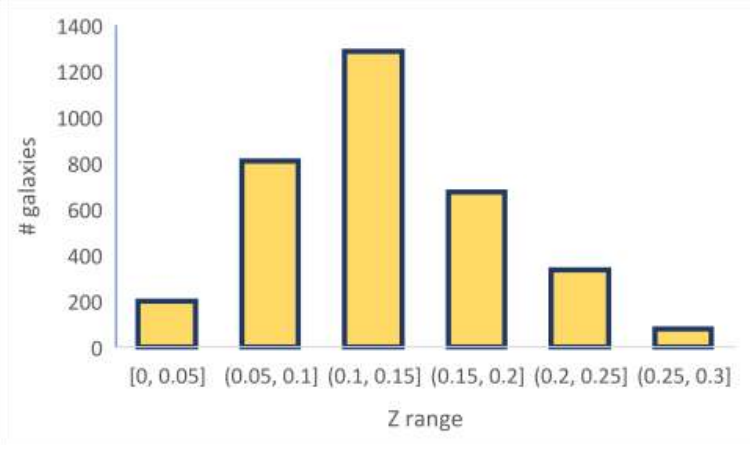}
\caption{The redshift distribution of the galaxies in the dataset of galaxies near the Southern Galactic pole.}
\label{redshift_range_decam}
\end{figure}

The entire dataset of annotated galaxies contains $\sim 8.07 \cdot 10^6$ galaxies, but because only galaxies with spectra in the 20$\times$20 field centered at the Galactic pole can be used, the dataset used here is reduced to 3,383 galaxies. Figure~\ref{redshift_range_decam} shows the redshift distribution of the galaxies. The dataset of annotated galaxies is available at  \url{https://people.cs.ksu.edu/~lshamir/data/zdifference/}.

Table~\ref{table_southern} shows the mean redshift of the galaxies that rotate in the same direction relative to the Milky Way and in the opposite direction relative to the Milky Way. Due to the perspective of the observer galaxies that are close to the Southern Galactic pole that rotate in the same direction relative to the Milky Way seem to rotate in the opposite direction compared to galaxies in the Northern Galactic pole that rotate in the same direction. 

\begin{table*}
\caption{The mean redshift of galaxies in 20$\times$20 field and the 10$\times$10 field centered at the Southern Galactic pole. The P values are the one-tailed Student t-test P values.}   
\label{table_southern}
\centering
\begin{tabular}{lcccccc}
\hline
Field  ($^o$)  & \# OMW  & \# MW  & $Z_{omw}$  & $Z_{mw}$  & $\Delta$z   & t-test p \\
\hline
10$\times$10     &   414   &  376   & 0.1270$\pm$0.0025   &  0.1352$\pm$0.0027  & -0.0082$\pm$0.0036 & 0.018 \\  
20$\times$20     &   1702 &  1681  & 0.1273$\pm$0.0014  &  0.1317$\pm$0.0013 &  -0.0044$\pm$0.0018 & 0.008 \\  
\hline
\end{tabular}
\end{table*}

As the table shows, the redshift differences are statistically significant in both fields, and increases when the galaxies are closer to the Galactic pole. These results are in good agreement with the results shown with galaxies located around the Northern Galactic pole. The table also shows that the mean redshift is higher compared to the mean redshift observed with SDSS. That difference can be expected due to the superior imaging power of DECam compared to SDSS, allowing DECam image galaxies at deeper redshifts. Unlike SDSS, where all redshifts are collected by the same instrument, the SIMBAD database collects redshifts from several different available sources, each source might use a different instrument. Still, it is reasonable to believe that the redshifts are distributed uniformly across the instruments that collected them, with no link between the rotation of direction and the specific instrument that provided the reddshift.

When combining the data from the Southern Galactic pole with the data from the Northern Galactic pole described in Section~\ref{data}, the mean z for galaxies that rotate in the same direction relative to the Milky Way is 0.11985$\pm$0.001, while galaxies that rotate in the opposite direction relative to the Milky Way have mean z of 0.11476$\pm$0.001. That provides a $\Delta z$ of 0.0051$\pm$0.0015, and the two-tailed Student t-test probability to have such difference by chance is $p\simeq$0.0003.

\section{Possible explanations and future experiments}
\label{explanation}

Recent puzzling observations such as the $H_0$ tension and large disk galaxies at high redshifts have been challenging cosmology. Explaining such observations require to assume that either the standard cosmological models are incomplete, or that the redshift as a distance model is incomplete. This study shows first direct observational evidence of bias in the redshift as a distance indicator. 

While the link between the redshift of the galaxy and its direction of rotation relative to the Milky Way is consistent, it is definitely difficult to provide an immediate trivial explanation to that link. The redshift of galaxies has a linear velocity component, but also a rotational velocity component. Since the rotational component is expected to be very small compared to the linear component, most models ignore the rotational component, and use the redshift base on the linear velocity component alone. The analysis shown here provides evidence that the rotational component of the redshift might not be negligible. Although the analysis does not provide a straightforward explanation to the anomaly, it should be reminded that the physics of galaxy rotation is still not fully understood. Explanations such as dark matter or MOND are the common explanations to the physics of galaxy rotation, but despite several decades of research there is no proven explanation to that physics. Due to the mysterious nature of the physics of galaxy rotation, a link between the galaxy rotation anomalies and the observation reported in this paper is one of the possible directions to explaining the observation.

Besides the theoretical aspects of the observation reported in this paper, redshifts are used in practice as the most common way to determine distances at cosmological scales. A possible systematic bias might therefore have impact on other studies that rely on redshift as a distance metrics. Although the bias is relatively small, it might have impact on experiments that use populations of galaxies to study the large-scale structure of the Universe. Examples of such studies that can potentially be affected by subtle redshift biases will be discussed later in this section.

The analysis done here is limited to galaxies at relatively low redshift of $z<0.3$. When the redshift of the galaxies is higher the velocity is also higher, and therefore the rotational velocity of the galaxies compared to the linear velocity is smaller. That can lead to a smaller change in the redshift relative to the linear velocity of the galaxy. To test that, an experiment can be done by imaging a relatively deep field at the galactic pole with space-based instruments such as HST or JWST. The spectra of galaxies in that field can provide information of the effect of the rotational velocity to galaxies at higher redshifts. 

Also, deeper and larger datasets of clear galaxies with spectra such as the data provided by the Dark Energy Spectroscopic Instrument (DESI) will allow a more detailed profiling of the observed anomaly. While  the observations do not explain directly the existence of early massive disk galaxies as revealed by JWST, they demonstrate that the current redshift model might be incomplete, and might need to be expanded to also include the rotational velocity of the galaxy to better estimate its distance. In that case, the existence of such galaxies can be explained without the need to modify the standard cosmological models.

While the bias can also be attributed to the algorithm that selects spectroscopic targets, it is difficult to think of how that algorithm could be affected by the direction of rotation relative to the Milky Way. Also, if the target selection algorithm has such unknown and complex biases, these biases are expected to be consistent throughout the sky, and is not expected to decrease when the angular distance of the galaxy from the Galactic pole gets larger, or flip when analyzing galaxies from the opposite side of the Galactic pole. The fact that two different telescopes show similar results further reduces the possibility that the results are driven by an unknown anomaly in the selection algorithm of the spectroscopic surveys, or another unexpected anomaly in the telescope system.


Another possible explanation to the observation is an unexpected anomaly in the geometry of the Universe or its large-scale structure. While a certain alignment in galaxy direction of rotation is expected \citep{d2022intrinsic,kraljic2020and}, if the observation reported in Section~\ref{results} reflect the real distribution of galaxies in the Universe, the scale of that structure that covers two hemispheres is far larger than any known supercluster of filament. If the redshifts represent the accurate distances of the galaxies, and is not affected by their rotational velocity, the galaxies form a cosmological-scale structure driven by the alignment in the direction of rotation of the galaxies, and peaks around the Galactic pole. That explanation assumes no anomaly in the physics of galaxy rotation, but it is aligned with cosmological models that shift from the standard model \citep{aluri2023observable}. As discussed also in \citep{shamir2022analysis,shamir2022analysis2,shamir2022large,shamir2022asymmetry,shamir2022possible,shamir2023large}, the observation of such large-scale structure that forms a cosmological-scale axis is aligned with alternative theories such as dipole cosmology \citep{ebrahimian2023towards,krishnan2022tilt,allahyari2023big,krishnan2023dipole,krishnan2023copernican}, or theories that assume a rotating universe \citep{godel1949example,ozsvath1962finite,godel2000rotating,chechin2016rotation,camp2021} such as Black Hole Cosmology \citep{pathria1972universe,stuckey1994observable,easson2001universe,seshavatharam2010physics,poplawski2010radial,christillin2014machian,dymnikova2019universes,chakrabarty2020toy,poplawski2021nonsingular,seshavatharam2022concepts,gaztanaga2022black,gaztanaga2022black2} which is also linked to holographic universe \citep{susskind1995world,bak2000holographic,bousso2002holographic,myung2005holographic,hu2006interacting,rinaldi2022matrix}. The presence of a cosmological-scale axis agrees also with the contention of ellipsoidal universe \citep{campanelli2006ellipsoidal,campanelli2007cosmic,gruppuso2007complete,campanelli2011cosmic,cea2014ellipsoidal}. In that case, the alignment of such hypothetical axis with the Galactic pole is a coincidence. The observation described in this paper can also be related to the theory of stationary Universe, which explains multiple anomalies but is challenged by the luminosity-distance relationship \citep{sanejouand2022framework}.

The results shown here might also provide an indication that the $H_0$ tension can be explained by the slight differences in the redshift. While $H_0$ anisotropy has been reported in the past \citep{javanmardi2015probing,krishnan2022hints,cowell2022potential,cowell2022potential2,aluri2023observable}, its nature is still unclear. But these studies are normally based on a limited number of galaxies with redshift that also host Ia supernovae. A higher number of galaxies that rotate in a certain direction can lead to a slight difference in the $H_0$, and therefore to $H_0$ anisotropy.  These reports can be linked to the observed anisotropy in the brightness of Ia supernova \citep{perivolaropoulos2023isotropy}. Explaining the $H_0$ tension might require new physics, which is not necessarily limited to physics that applies to the early Universe alone \citep{vagnozzi2023seven}.  

Differences in the redshift that are based on the rotational velocity of the galaxies relative to the Milky Way can explain the $H_0$ anisotropy, and potentially also the $H_0$ tension. If the rotational velocity of Ia supernovae and their host galaxies relative to the Milky Way affect their estimated distance, when the rotational velocity relative to the Milky Way is normalized, the $H_0$ tension is expected to be resolved. As demonstrated in \citep{mcadam2023asymmetry}, when limiting the {\it SH0ES} collection \citep{refId0} of Ia supernovae to galaxies that rotate in the same direction as the Milky Way, the computed $H_0$ drops to $\sim$69.05 $km/s Mpc^{-1}$, which is closer to the $H_0$ determined by the CMB, and within statistical error from it. When using just galaxies that rotate in the opposite direction relative to the Milky Way, the $H_0$ does not drop, but instead increases to $\sim$74.2 $km/s Mpc^{-1}$ to further increase the $H_0$ tension \citep{mcadam2023asymmetry}. While the analysis was done with the relatively small collection of {\it SH0ES}, it suggests that the distance indicators might respond to the rotational velocity of the observed objects compared to the rotational velocity of the Milky Way. This explanation agrees with the contention that the explaining the $H_0$ tension might require new physics, which is not necessarily limited to physics that applies to the early Universe alone \citep{vagnozzi2023seven}.

The observed $\Delta z$ between galaxies with opposite rotational velocities as shown here is between around 0.0065 to 0.012. If that difference is due to the rotational velocity, that difference corresponds to velocity of between roughly 2,000 to 3,600 km$\cdot$s$^{-1}$. That is about 5 to 8 times the rotational velocity of the Milky Way compared to the observed galaxies, which is $2\cdot220=\sim 440$ km$\cdot$s$^{-1}$, assuming that the observed galaxies have the same rotational velocity as the Milky Way. That velocity difference is in good agreement with the velocity difference predicted in \citep{shamir2020asymmetry} by using analysis of the photometric differences between galaxies rotating with or against the rotational velocity of the Milky Way. That analysis was based on the expected and observed differences in the total flux of galaxies that rotate in the same direction relative to the Milky Way and the flux of galaxies that rotate in the opposite direction. Based on the expected flux difference due to the Doppler shift driven by the rotational velocity as shown in \citep{loeb2003periodic}, it was predicted that light emitted from the observed galaxies agrees with rotational velocity that is 5-10 times faster than the rotational velocity of the Milky Way \citep{shamir2020asymmetry,mcadam2023asymmetry}. These predictions are close to the results of comparing the redshift as done here.

The observed redshift difference, if indeed driven by the rotational velocity of the Milky Way and the observed galaxies, corresponds to a much higher rotational velocity than the $\sim$220 km$\cdot$s$^{-1}$ rotational velocity of the Milky Way. Such consistent bias in the redshift can be linked to other unexplained anomalies that challenge cosmology, and related to distances at cosmological scale. The physics of galaxy rotation is one of the most puzzling and provocative phenomena in nature, and despite over a century of research it is still not fully understood \citep{opik1922estimate,babcock1939rotation,oort1940some,rubin1970rotation,rubin1978extended,rubin1980rotational,rubin1985rotation,sanders1990mass,rubin2000,mannheim2006alternatives,kroupa2012dark,kroupa2012failures,kroupa2015galaxies,arun2017dark,akerib2017,bertone2018new,aprile2018,skordis2019gravitational,sivaram2020mond,hofmeister2020debate,byrd2021spiral,haslbauer2022has}.  Due to the unexplained tensions in cosmology, the unknown physics of galaxy rotation should be considered as a factor that can be associated with these tensions and explain them.

\section{Conclusion}
\label{conclusion}

Recent puzzling observations such as the $H_0$ tension and large disk galaxies at high redshifts have been challenging cosmology. Explaining such observations require to assume that either the standard cosmological models are incomplete, or that the redshift as a distance model is incomplete. This study shows first direct observational evidence of bias in the redshift as a distance indicator. The bias is relatively small, and was observed in galaxies with relatively low redshift. Studies that are based on a relatively small number of galaxies such the anisotropy of $H_0$ might be affected by the distribution of galaxies that rotate in opposite direction. If the direction of rotation of the galaxies are not normalized, a slightly higher prevalence of galaxies that rotate in a certain direction can lead to small but consistent bias.

There is no immediate proven physical explanation to the difference between the redshifts of galaxies that rotate with or against the direction of rotation of the Milky Way. While a certain difference is expected, the magnitude of the difference is expected to be small given the rotational velocity of the Milky Way. But as discussed above, the physics of galaxy rotation is not fully understood. Namely, the rotational velocity of galaxies cannot be explained without assumptions such as dark matter or MOND. It is therefore possible that the mysterious physics of galaxy rotation affects the light emitted by galaxies in a way that is not proportionate to their rotation velocity.

Future experiments will include using larger redshift survey such as DESI to profile the redshift differences using a far higher number of galaxies with redshift. Other experiments can be based on space-based instruments pointing at the galactic poles to study the redshift difference at higher redshifts and the earlier Universe.

\section*{Data availability}

All data used in this paper are available publicly. The coordinates, redshift, and direction of rotation of galaxies of the datasets of galaxies from the Northern and Southern Galactic pole are available at \url{https://people.cs.ksu.edu/~lshamir/data/zdifference/}. The information of galaxies annotated by {\it SPARCFIRE} are available at \url{https://people.cs.ksu.edu/~lshamir/data/sparcfire/}. {\it Galaxy Zoo} data are available through SDSS CAS \url{http://casjobs.sdss.org/CasJobs/default.aspx}.

\section*{Acknowledgments}

I would like to thank the three anonymous reviewers and the associate editor for the helpful comments. This study was supported in part by NSF grant 2148878.



\bibliographystyle{apalike}
\bibliography{main}

\end{document}